\documentclass[8.5pt,twoside,twocolumn]{article}
\usepackage[super,sort&compress,comma]{natbib}
\usepackage{mhchem}
\usepackage{times,mathptmx}
\usepackage{sectsty}

\usepackage{graphicx} 
\usepackage[format=plain,justification=raggedright,singlelinecheck=false,font=small,labelfont=bf,labelsep=space]{caption}
\usepackage{fancyhdr}
\usepackage{color}
\usepackage{hyperref}
\begin{document}

\thispagestyle{plain}
\fancypagestyle{plain}{
\renewcommand{\headrulewidth}{1pt}}
\renewcommand{\thefootnote}{\fnsymbol{footnote}}
\renewcommand\footnoterule{\vspace*{1pt}%
\hrule width 3.4in height 0.4pt \vspace*{5pt}}
\setcounter{secnumdepth}{5}

\makeatletter
\def\subsubsection{\@startsection{subsubsection}{3}{10pt}{-1.25ex plus -1ex minus -.1ex}{0ex plus 0ex}{\normalsize\bf}}
\def\paragraph{\@startsection{paragraph}{4}{10pt}{-1.25ex plus -1ex minus -.1ex}{0ex plus 0ex}{\normalsize\textit}}
\renewcommand\@biblabel[1]{#1}
\renewcommand\@makefntext[1]%
{\noindent\makebox[0pt][r]{\@thefnmark\,}#1}
\makeatother
\renewcommand{\figurename}{\small{Fig.}~}
\sectionfont{\large}
\subsectionfont{\normalsize}


\twocolumn[
  \begin{@twocolumnfalse}

\noindent\LARGE{\textbf{Predicting genetic interactions from Boolean models of biological networks}}
\vspace{0.6cm}

\noindent\large{\textbf{Laurence Calzone,\textit{$^{a,b,c,d}$} Emmanuel Barillot,\textit{$^{a,b,c,e}$} and
Andrei Zinovyev\textit{$^{a,b,c,f}$}}}\vspace{0.5cm}


\noindent \normalsize{
{\bf Abstract:} Genetic interaction can be defined as a deviation of the phenotypic quantitative effect of a double gene mutation from the effect predicted from single mutations using a simple (e.g., multiplicative or linear additive) statistical model. Experimentally characterized genetic interaction networks in model organisms provide important insights into relationships between different biological functions. We describe a computational methodology allowing to systematically and quantitatively characterize a Boolean mathematical model of a biological network in terms of genetic interactions between all loss of function and gain of function mutations with respect to all model phenotypes or outputs. We use the probabilistic framework defined in MaBoSS software, based on continuous time Markov chains and stochastic simulations. In addition, we suggest several computational tools for studying the distribution of double mutants in the space of model phenotype probabilities. We demonstrate this methodology on three published models for each of which we derive the genetic interaction networks and analyze their properties. We classify the obtained interactions according to their class of epistasis, dependence on the chosen initial conditions and phenotype. The use of this methodology for validating mathematical models from experimental data and designing new experiments is discussed.$^\dag$
}
\vspace{0.5cm}
 \end{@twocolumnfalse}
]

\section{Introduction}

Genetic interaction is defined as a phenomenon by which the effect of a double gene mutation cannot be predicted from the effect of single mutations using a simple (such as additive or multiplicative) statistical model \cite{Fisher1918, Mani2008, Segre2005}. The strength of the interaction can be characterized by an epistatic score, which is, in the case of purely deleterious mutations, negative for synergistic interactions (when the phenotype of a double mutation is significantly stronger than the expected combined effect of two independent single mutations), and positive for alleviating interactions (when the combined effect is weaker). Examples of synergistic interactions are synthetic lethality and synthetic sickness (in the case of survival-related phenotype) or synthetic enhancement of a phenotype \cite{Guarente1993, Boone2007}. An example of strong alleviating interaction is the suppression of an effect of one mutation by a second mutation (in classical genetics, such interactions were historically defined as ``epistatic"). Genetic interactions in the general case of both beneficial and deleterious mutations can be classified into 9 groups according to various inequality relations between the effects of single and double mutants \cite{Drees2005}.

Genetic interaction networks provide important insights into relations between different biological functions \cite{Stern2009}. Knowledge of genetic interactions with respect to a disease phenotype can provide important hints on personalized treatment strategy, in particular, in cancer \cite{Kaelin2005, Barillot2012, Nijman2013}. This knowledge is currently obtained by costly high-throughput screening techniques based on knocking-out or knocking-down genes (using siRNA or shRNA) in model organisms, such as yeast \cite{Costanzo2010, Tong2004}, worm \cite{Bussey2006}, mouse \cite{Einav2005} or human cells \cite{Nijman2011}. Experimentally, one can measure both synthetic and synthetic dosage interactions \cite{Paul2014}. Establishing single genetic interactions can be a result of long and tedious work, in the case of phenotypes that are complex and difficult to observe such as metastasis \cite{Chanrion2014}.

Computational approaches have been used in order to derive genetic interactions from dynamical mathematical models or by using machine learning approaches. One of the earliest attempts to characterize the genetic networks of the genes involved in metabolism was done using flux balance analysis framework applied to a genome-wide reconstruction of yeast metabolic network\cite{Segre2005}. In this work, the quantitative epistatic measure was introduced to characterize the genetic interactions as a difference between the observed effect of a double mutant and the multiplicative model prediction from the effect of two single mutation effects. It was noted that the distribution of the epistatic measure is tri-modal and that the interactions between functional modules have a tendency for monochromaticity, i.e., having the same dominant sign for between-module interactions. In a recent paper, a similar approach was applied to characterize genetic interactions with respect to multiple metabolism-related phenotypes\cite{Snitkin2011}.

There have been many attempts to apply machine learning approach for predicting genetic interactions from a subset of known interactions\cite{Steen2012, Boucher2013}. For instance, in yeast, the structure of physical interaction networks was combined with co-expression networks; data on protein classification was used for predicting genetic interactions\cite{Wong2004}. In worm, identical anatomical expression and microarray co-expression, phenotype proximity, Gene Ontology annotation and presence of interlogs were the parameters used for fitting the logistic regression in order to score genetic interactions\cite{Zhong2006}. Decision tree-based approaches trained on the structure of protein-protein interaction and co-expression networks in both yeast and worm were also used\cite{Chipman2009}. Short polypeptide cluster detection was utilized to predict synthetic lethal interactions between genes in yeast\cite{Zhang2012}. Still in yeast, evolutionary approaches and the notion of functional asymmetry allowed prediction of negative genetic interactions between protein complex components\cite{Lu2013}. There are very few examples of computational predictions of genetic interactions in human, one of them used gene expression analysis to predict synthetic lethal partners of TP53 gene\cite{Wang2013}. The main problem of most of machine learning approaches is the absence of {\it bona fide} negative example (absence of interaction) set for training, which is usually needed for a successful application of automated classification methods \cite{Boucher2013}. Nevertheless, it was shown that machine learning methods are able to predict genetic interactions significantly better than random choice of a gene pair.

The knowledge about molecular mechanisms involved in a biological phenomenon that one wishes to study can be represented as a network of interacting entities\cite{Novere2009}. Depending on the network type, the translation into a mathematical model can be done using an appropriate formalism (ordinary or partial differential equations, logical, rule-based modeling, etc.). These mathematical models can predict the effect of a perturbation, intrinsic or extrinsic, and anticipate the response of a drug, for instance. Boolean (or, more generally, logical) modeling focuses on how the influences of regulatory molecules combine to control the expression or activity of each molecular entity - or process - composing the regulatory network. In a purely Boolean framework, each variable of the model can only take two values: 0 or 1 (absent/inactive or present/active). In our studies, we found that Boolean formalism represents a convenient mean of abstraction for modeling cellular biochemistry dynamics and verifying that the topology of the networks representing the studied phenomena fits the experimentally-observed effects of loss or gain of function mutations on a phenotype. So far, there was no attempt to systematically predict genetic interactions using Boolean models of biological mechanisms.

An important remark should be made with respect to any attempt to predict the genetic interactions computationally. Genetic interactions, being functional rather than physical, can strongly depend on the choice of both the phenotype (or model read-out) and the set of initial conditions used for model simulations. Therefore, genetic interactions can be classified as occurring with respect to single versus multiple phenotypes, and dependent versus independent on initial conditions. With the mathematical model of metabolism in yeast, it was shown that genetic interactions synergistic with respect to one phenotype can become alleviating with respect to another one\cite{Snitkin2011}. Similarly, depending on the set of initial conditions (accounting for homeostatic, physiological, nutrient-deprived, etc. conditions), some phenotypes represented in the model can never be reached or the simulations can lead to a different output with the same set of inputs. For example, in a model of cell fate decision process in response to TNF (or Fas) ligand activation signal, the cell response showed to be either survival or cell death (non-apoptotic and apoptotic with a higher probability for necrotic phenotype though) depending on the activity of some nodes of the model \cite{Calzone2010}. In a model describing the kinetics of the restriction point, if the G1 cell cycle phase cyclin, Cyclin D1 (CycD in the model), is initially active (corresponding to presence of growth factors), the cell enters the cycle, otherwise, it stays stuck in G1 arrest \cite{Novak2004, Faure2006}.

In this manuscript, we suggest a quantitative methodology to convert a logical model of a regulatory network into a genetic interaction network, defined with respect to a chosen model phenotype (which can be any phenotype and not only survival-related as it is often the case). The methodology is based on using the formalism of continuous time Markov chains implemented in MaBoSS software \cite{Stoll2012}. Using published models, we applied our method to derive several genetic interaction networks for the genes that compose these models. We analyze genetic network properties and show that they possess many features of experimentally-measured genetic networks. The derived genetic interactions reflect the functional properties of the mathematical models studied, so we briefly compare these predicted functional relations using available databases.

\footnotetext{\dag~Electronic Supplementary Information (ESI) available: http://maboss.curie.fr/gins. See DOI: 10.1039/b000000x/}

\footnotetext{\textit{$^{a}$~Institut Curie, 26 rue d'Ulm, Paris, France}}
\footnotetext{\textit{$^{b}$~INSERM U900. Paris, France. }}
\footnotetext{\textit{$^{c}$~Mines ParisTech, Fontainbleau, France}}
\footnotetext{\textit{$^{d}$~E-mail: Laurence.Calzone@curie.fr}}
\footnotetext{\textit{$^{e}$~E-mail: Emmanuel.Barillot@curie.fr}}
\footnotetext{\textit{$^{f}$~E-mail: Andrei.Zinovyev@curie.fr}}



\section{Methods and data}

\subsection{Models used in this study}

Three published models were selected for testing the method. The models correspond to signalling pathways involved in cancer with the focus on: the MAPK pathway\cite{Grieco2013} describing the crosstalk between the three mitogen-activated protein kinases: ERK, p38 and JNK, and their role in apoptosis and proliferation balance; the cell cycle with the focus on the biochemical processes regulating the restriction point\cite{Novak2004, Faure2006}; and cell fate decision between survival and death in response to extrinsic signals such as death receptor activation\cite{Calzone2010,Zinovyev2012}.

For each of the model, we provide the models in both GINsim\cite{Naldi2009} and MaBoSS\cite{Stoll2012} formats. Several genetic interaction networks (GINs) per model can be constructed corresponding to different initial conditions and to the chosen phenotype. They can be found as separate Cytoscape sessions in Supplementary materials (Supp\_Mat\_GINs).

\subsection{Computing phenotype probabilities}

For each model, we computed the probability of reaching model phenotypes for all possible single and double mutants (resulting either from gain of function - modelled as fixing the corresponding node value to 1, and referred to as "overexpression" or "oe", or from loss of function - modelled as fixing the node value to 0 and referred to as "deletion" or "ko"). For these computations, we used both MaBoSS software and a set of scripts for processing the MaBoSS configuration and output files, implemented into BiNoM Cytoscape plugin \cite{Zinovyev2008, Bonnet2013, Bonnet2013a}. MaBoSS is a C++ software designed for simulating continuous/discrete time Markov processes, defined on the state transition graph representing the dynamics of a Boolean network. MaBoSS allows the modeller to associate different rates up and rates down to each variable of the model when the dynamics is known, enabling to account for different time scales of the processes described by the model. Given some initial conditions, MaBoSS computes time trajectories by applying Monte Carlo kinetics algorithm (Details and examples can be found at: {\small \url{http://maboss.curie.fr}}). More precisely, probabilities to reach a phenotype are computed as the probability for the variable associated to the phenotype to have the value 1, by simulating random walks on the probabilistic state transition graph. {The parameters for the stochastic simulations (number of runs, initial conditions, maximum time, etc.) are configured for each simulation.} The read-out can be a variable representing the phenotype, a variable representing a protein or gene, or a combination of them. The probabilities for the selected outputs are reported for each mutant based on predefined initial conditions (which can be all random). Since a state in the state transition graph can combine activation of several phenotype variables, some phenotype probabilities appear to be ``mixed" or coupled. It is particularly the case for cyclic attractors. For the cell fate model, we investigated the effect of the choice of the initial conditions (``random" versus ``physiological") on the final phenotype probability distribution. The result of the simulations is stored in a simple table, containing the complete set of mutants characterized by probabilities of all pure and mixed model phenotypes (in Supp\_Mat\_Models and Supp\_Mat\_GINs).

\subsection{Quantifying epistasis in double mutants}

\subsubsection{Definition of epistasis measures}.

The results of double mutant simulations were used to quantify the level of epistasis between two model gene defects $A$ and $B$ with respect to a particular phenotype $\phi$.
We define the normalized ``fitness" of a mutation (or combination of mutations) $X$ with respect to a phenotype $\phi$ as the ratio between the probability of the phenotype in the mutant $X$ and the wild-type models.

\begin{equation}
f^X_{\phi} = \frac{p^X_{\phi}}{p^{wt}_{\phi}}.
\end{equation}

To fully characterize a genetic interaction, one should be able to characterize its strength and type. We defined the strength of the interaction as a deviation of the fitness of the double mutant from one of the four simplest statistical models frequently used in this context: additive, logarithmic, multiplicative and min \cite{}, i.e.,

\begin{equation}
\epsilon_{\phi}(A,B)=f^{AB}_{\phi}-\psi(f^A_{\phi},f^B_{\phi}).
\label{epistasis_eps}
\end{equation}

\noindent where $f^A_{\phi}$ and $f^B_{\phi}$ are phenotype $\phi$ fitness values of single gene defects, $f^{AB}_{\phi}$ is the phenotype $\phi$ fitness of the double mutant, and $\psi(x,y)$ is one of the four functions:

\begin{equation}
\begin{array}{lcl}
\psi^{ADD}(x,y) = x+y & (additive) \\
\psi^{LOG}(x,y) = log_2((2^x-1)(2^y-1)+1) & (log) \\
\psi^{MLT}(x,y) = xy & (multiplicative) \\
\psi^{MIN}(x,y) = \min(x,y) & (min) \\
 \end{array}
\label{psi_def}
\end{equation}

To choose the best definition of $\psi(x,y)$, the Pearson correlation coefficient was computed between the fitness values observed in all double mutants and estimated by the null model. The null model with maximal linear correlation was chosen:

\begin{equation}
\begin{split}
\psi(x,y) = \arg \max_{\psi^{(i)}} corr(\psi^{(i)}(f^A_{\phi},f^B_{\phi}),f^{AB}_{\phi}), \\
i = {ADD, MLT, LOG, MIN}.
\end{split}
\end{equation}

Note that the best definition of $\psi$ can vary from model to model, from phenotype to phenotype, and even for different choices of initial conditions. Our simulations show that $\psi^{LOG}$ performs uniformly optimal or close to optimal in most of the simulations, having also advantage of not producing biased distributions of $\epsilon$ (see next section).

\subsubsection{Removing bias in the distribution of epistatic measure values}.

\begin{figure*}
\centering
  \includegraphics[width=16cm]{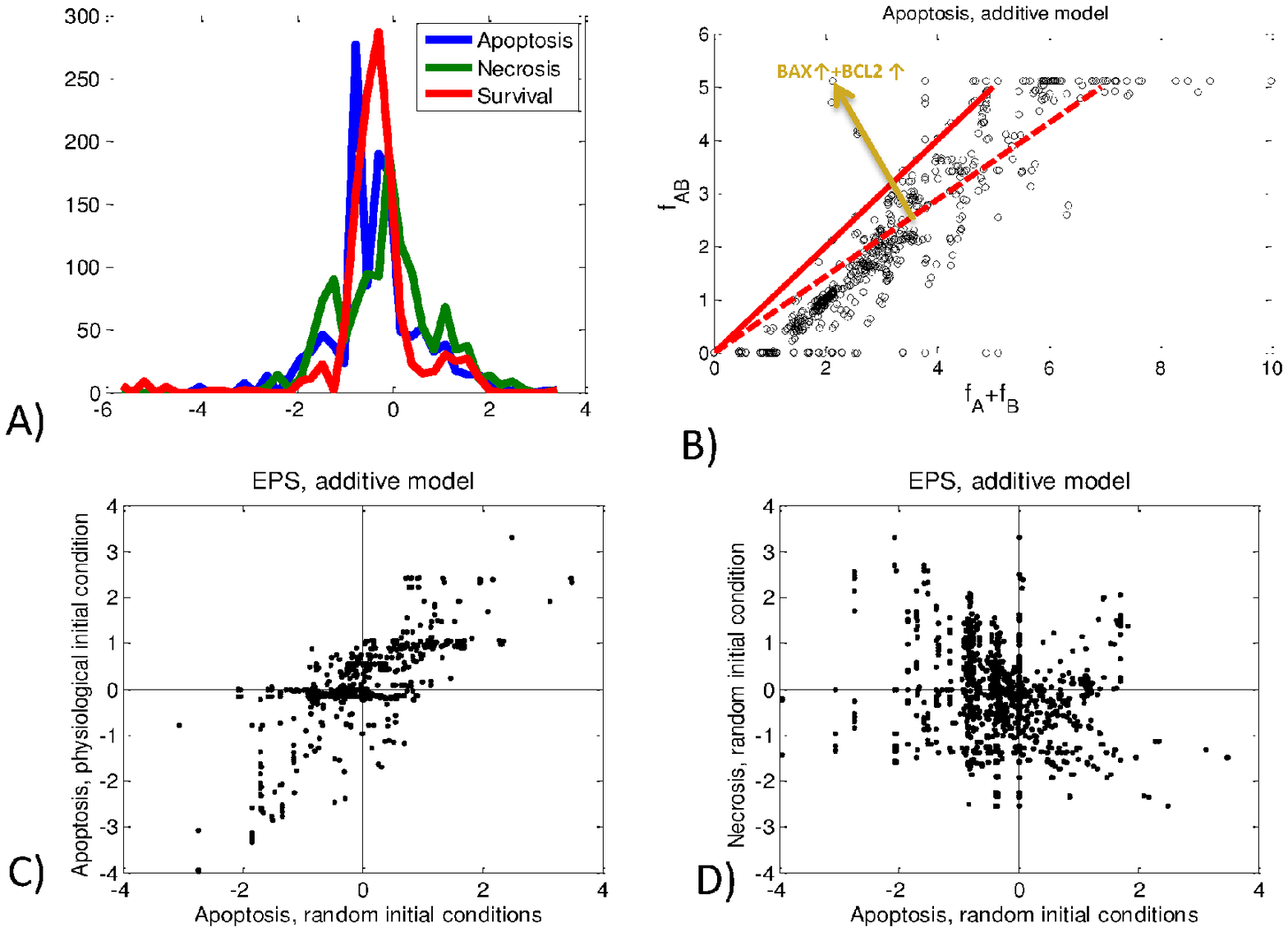}
  \caption{Illustrating epistasis measures for cell fate decision model. A) distribution of $\epsilon$ values for three phenotypes. B) Additive model of epistasis, solid line shows uncorrected additive null model and dashed line shows the corrected model; an arrow shows a particular double mutant BAX+/BCL2+, for which the combined effect is stronger than expected by the null model (example of single-nonmonotonic genetic interaction, $A<WT<B<AB$); the length of the arrow equals to $\epsilon(BAX+/BCL2+)$ in this case. C) comparison between $\epsilon$ values for the case of random initial conditions and the physiological initial condition. D) comparison between $\epsilon$ values for two different cell death phenotypes. }
  \label{cellfate_numbers}
\end{figure*}

After computing the distribution of epistatic measures, it can be observed that the peak of the distribution is shifted towards non-zero epistasis. This can be considered as a bias in estimating the null multiplicative model for quantifying the epistasis measure (\ref{epistasis_eps}). In our experiments, it was corrected by linear fitting of the observed value $y=f^{AB}$ to the null model $x=\psi(f^A,f^B)$ (see Figure~\ref{cellfate_numbers}B). Then the epistatic measure is defined as:

\begin{equation}
\epsilon^{(corrected)}_{\phi}(A,B)= f^{AB}_{\phi}-\alpha\psi(f^A_{\phi},f^B_{\phi}),
\label{epistasis_corrected}
\end{equation}

\noindent where $\alpha$ is the slope coefficient in the best linear fit estimation $\sum||y-\alpha x||^2 \to \min$. Further in the text, we refer to $\epsilon^{(corrected)}$ as $\epsilon$ unless explicitly specified.

\subsubsection{Choosing the threshold for defining the set of genetic interactions}.

The distribution of the epistasis measures $\epsilon$ is asymmetric in many examples. Therefore, we set a threshold separately for positive and negative part of epistastis measure distribution (Figure~\ref{cellfate_numbers}A) as a multiplier of one-tailed standard deviations. Those genetic interactions whose strength are above $k$ one-tailed standard deviations are selected, where $k$ is a real number parameter (typically, $k=2$ as a moderately stringent selection criterion).

\subsubsection{Defining the type of genetic interaction}.

Since in model simulations, one can have both deleterious $f^X<1$, neutral $f^X\approx 1$ and beneficial $f^X>1$ mutations $X$ with respect to a phenotype $\phi$, multiple possibilities arise for relations between four numbers $f^A$, $f^B$, $f^{AB}$ and $f^{WT}=1$ which cannot be simply grouped into alleviating and aggravating, as in the simplest case of pure deleterious mutations. We classified gene interactions using the existing approach\cite{Drees2005}, according to 75 possible inequalities between these four numbers which are further grouped into 9 genetic interaction classes: ``suppressive", ``epistatic", ``conditional", ``single-nonmonotonic", ``additive", ``double-nonmonotonic", ``non-interactive", ``synthetic", ``asynthetic". The first 4 classes in this list can be characterized by a direction of the interaction, i.e., mutation $A$ is epistatic to $B$ means that the effect of $A$ completely cancels the effect of $B$ (and both effects are different from the wild-type), and not the opposite ($A\rightarrow B$). Note that the directed genetic interaction maps the causal effects in opposite direction (e.g., mutations in downstream effectors of a phenotype can mask more upstream mutations).

To define inequalities, we introduced a threshold for distinguishing different values of fitness $f$, i.e., we consider two values of fitness $f^A$ and $f^B$ equal, if $|f^A-f^B|<\delta$, where we typically choose $\delta=0.2$.

For example, one of the most prevalent interactions in our simulations is the ``epistatic" (in the sense of the classical definition of the notion ``epistasis") interaction which corresponds to inequalities $B<WT<A=AB$ (denoting $f^B<f^{WT}=1<f^A=f^{AB}$) or $A=AB<WT<B$ meaning that the effects of single mutants are opposite with respect to the wild-type (one is deleterious and another is beneficial) and the effect of the double mutant is equal to one of the single mutants (one single mutant ``wins"). Another interesting example is ``synthetic" interaction type which can correspond to the inequality $AB<WT=A=B$ (classical ``synthetic sickness") or to $WT=A=B<AB$ (``synthetic enhancement").

\begin{figure*}
\centering
  \includegraphics[width=15cm]{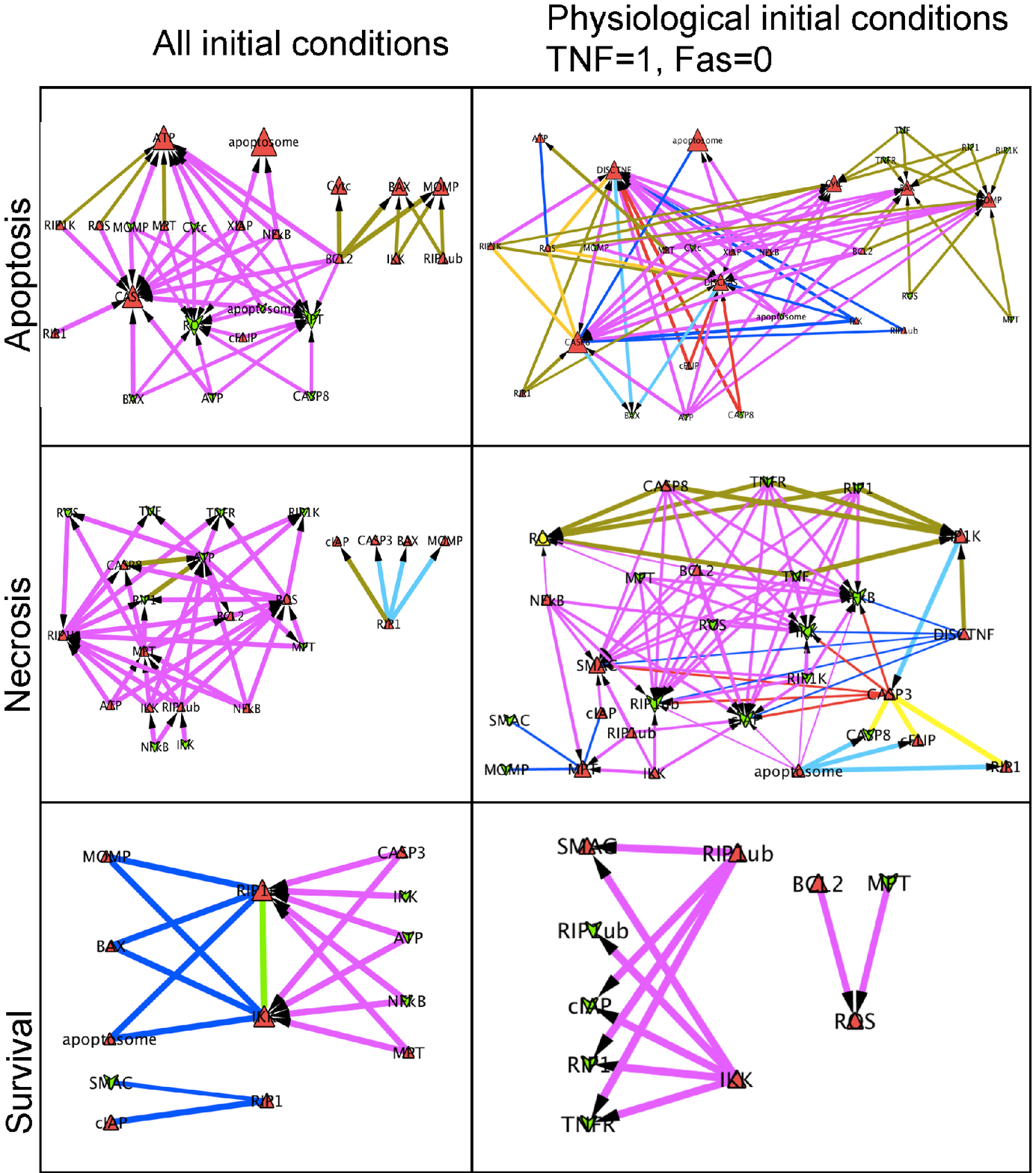}
  \caption{Genetic interaction networks computed for cell fate decision model, with random and physiological initial conditions and for the three considered phenotypes: apoptosis, necrosis and survival}
  \label{cellfate_gin}
\end{figure*}

Some interaction types are counter-intuitive such as ``single-nonmonotonic" which can correspond to the inequality $A<WT<B<AB$, when a combination of deleterious and beneficial mutations lead to enhancement of the phenotype stronger than the beneficial mutation alone. It was shown that these interactions are observed in real data\cite{Drees2005}, and they are also observed in some of our simulations (see Figure~\ref{cellfate_gin}).

\subsubsection{Visualizing genetic interaction network using Cytoscape}.

\begin{figure}
\centering
  \includegraphics[width=8cm]{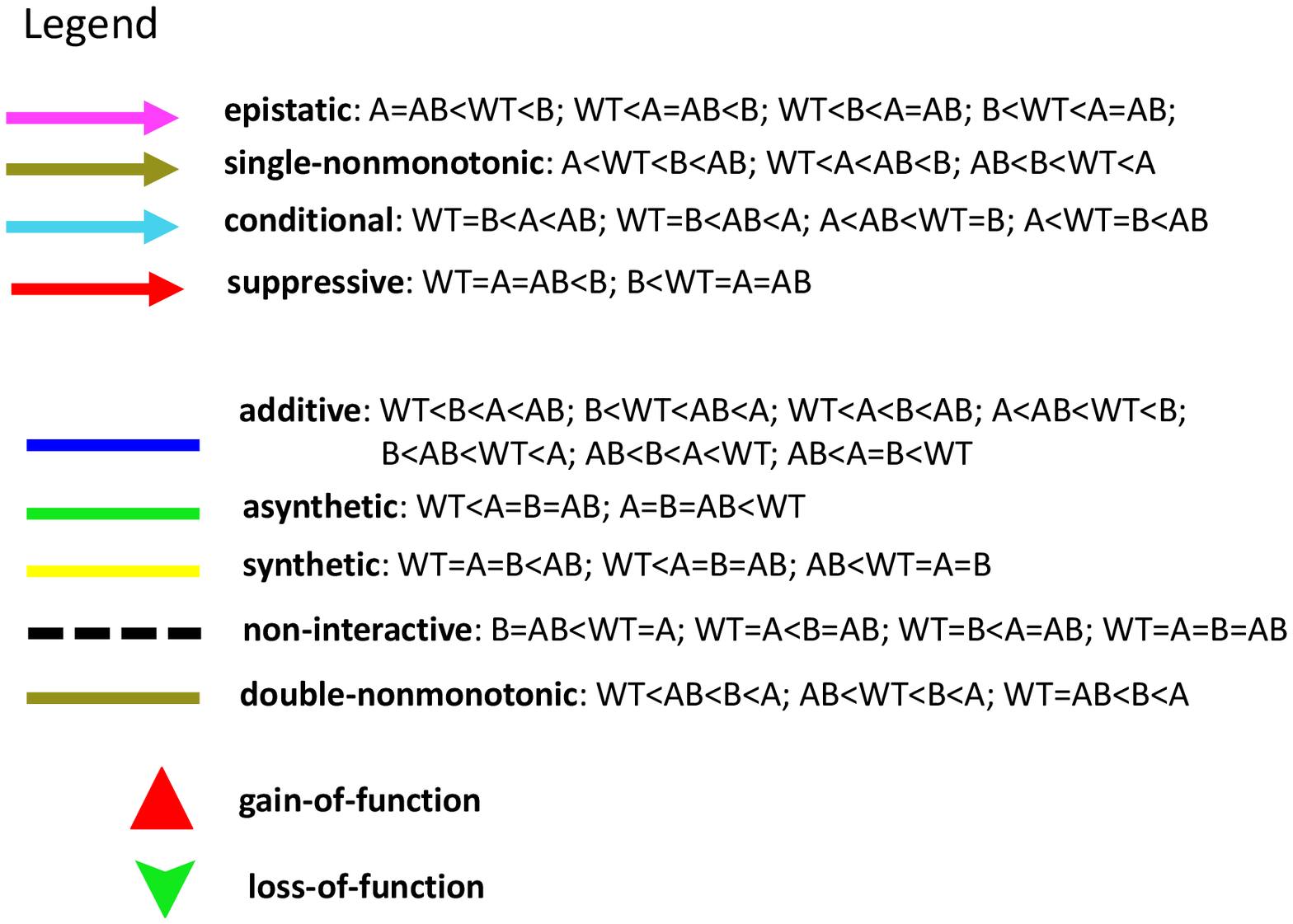}
  \caption{Colour code for the genetic interaction networks. The name of the interaction and the colour code is in accordance with \cite{Drees2005}. Only the rules found in our analyses of the three models are indicated for each interaction}
  \label{legend}
\end{figure}

The selected genetic interactions are visualized in Cytoscape\cite{Cline2007} (see example with Figure~\ref{MAPK_gin} and Figure~\ref{cellfate_gin}). The visual mapping chosen distinguishes, by colour and shape, loss of function and gain of function single mutants. Size of the nodes reflects the effect on the phenotype of a single mutant, and the width of the edge, the epistatic effect strength of the corresponding double mutant. Colouring edges denotes their types, using the colour schema suggested before\cite{Drees2005} (see Figure~\ref{legend} for definition of the visualization style).

\subsubsection{Using non-linear principal component analysis for mapping double mutant distribution in the space of phenotype probabilities}

\begin{figure}
\centering
  \includegraphics[width=8cm]{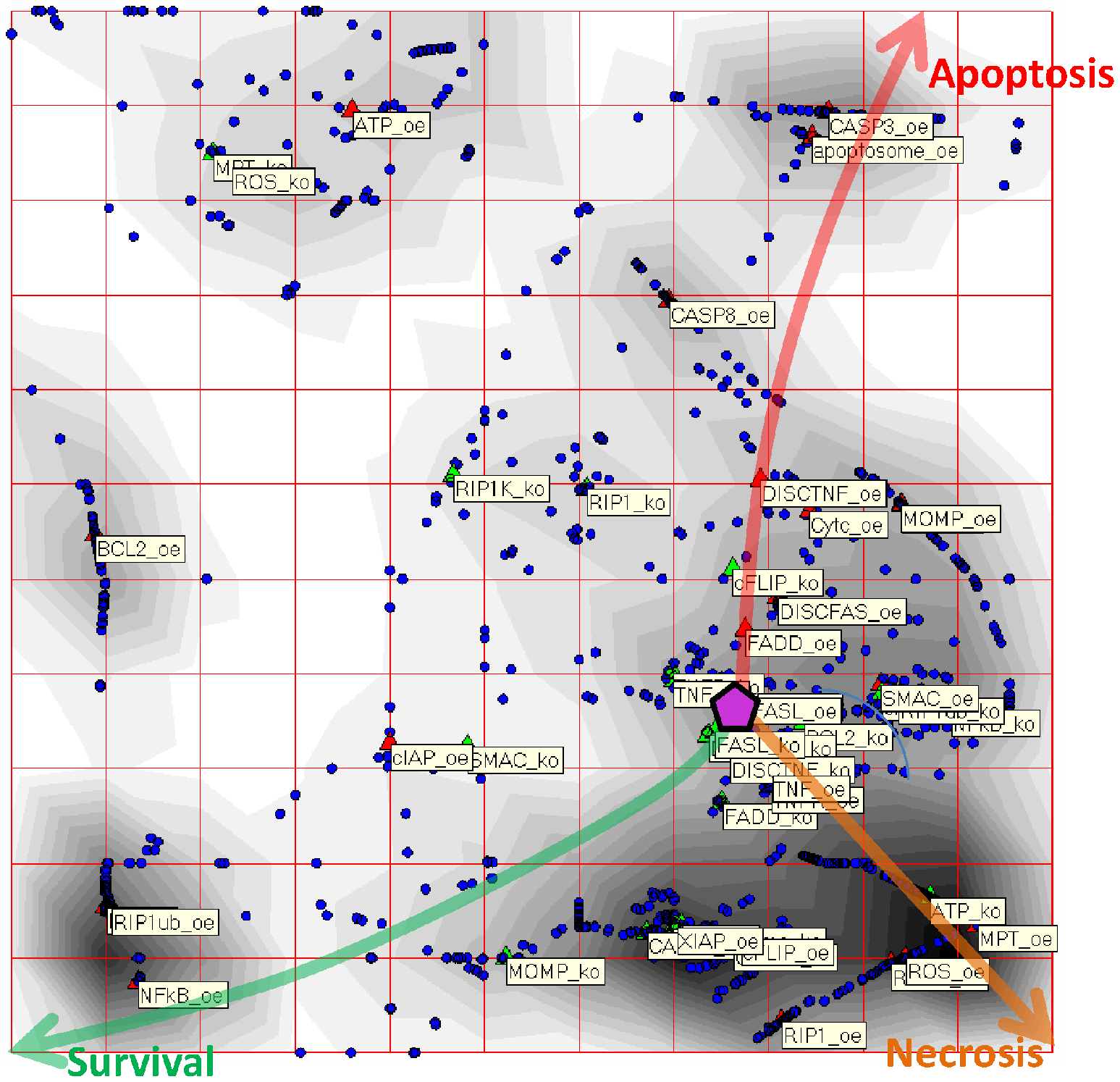}
  \caption{Application of non-linear principal manifold analysis for visualizing the distribution of double mutants in the space of phenotype probabilities. The figure shows projection of phenotype probabilities from multi-dimensional space onto the 2D space of internal coordinates of the non-linear principal manifold. Each point corresponds to a mutant. A big violet pentagon coresponds to the wild-type model, triangles to single-element mutant model and circles to double mutants. Gradients of increase of the model phenotypes probabilities are shown by curved arrows. The gray color in the background visualizes local density of the projections onto the map, allowing to perform cluster analysis visually.}
  \label{cellfate_elmap}
\end{figure}

The non-linear principal manifolds were constructed for the distribution of all single and double mutants of the model in the space of computed model phenotype probabilities, using elastic maps method \cite{Gorban2001,Gorban2008Principal,Gorban2010} and ViDaExpert software\cite{GorbanVidaExpert2014}. For computation, only the mixed phenotypes with a probability expectation over the whole set of double mutants with more than 1\% were selected. This results in sets of double mutants in multi-dimensional space for which principal manifolds were computed (see Figure~\ref{cellfate_elmap}).

\section{Results and discussion}

The three Boolean models were downloaded either from The Cell Collective database\cite{Helikar2012} or from GINsim database\cite{Naldi2009}. The stable state analysis was done in GINsim software. The models were then exported in MaBoSS for simulations. Finally, we used some scripts embedded into BiNoM cytoscape plugin to automatically compute probabilities for all single and double mutants (including both gain of function and loss of function mutants for all components of each model) and visualize the results of paired interactions as genetic interaction networks. A thorough description of each model is given in supplementary materials (SuppMat\_description\_models) along with the Cytoscape sessions for each model, each phenotype and different initial conditions for one of the models (SuppMat\_GINs).

\subsection{Cell fate decision model}

Figure~\ref{cellfate_gin} shows the genetic interaction networks computed with respect to three different phenotypes (survival, apoptosis, non-apoptotic cell death referred to as necrosis for short)\cite{Calzone2010}.

The general shape of the epistatic measure distribution exhibit tri-modality (Figure~\ref{cellfate_numbers}A), as it was previously observed in another modeling framework\cite{Segre2005}. The cell fate decision GINs and the distributions of $\epsilon$ show that the networks computed for different phenotypes are less similar than the networks computed for the same phenotype but with different initial conditions (Figures \ref{cellfate_gin} and \ref{cellfate_numbers}C,D, with the legend for GINs given in Figure~\ref{legend}). Similar conclusions were made for most of the constructed GINs in this study.

For the physiological initial conditions with TNF=1, some gene alterations (and, by extension, some pathways) appear to be more important than when all initial conditions are considered. Indeed, some of these interactions are lost in the numerous genetic interactions when considering all initial conditions. It is particularly evident for the survival phenotype. Overexpressing any gene from the survival pathway, which is described in a linear manner in this model is enough to favour or even force the survival phenotype. When taking in account all possible inputs, other pathways can help reach the survival phenotype: the additive effect of both RIP1 and cIAP gain of function would be equivalent to forcing RIP1ub.
Single-nonmonotonic interactions are found numerous in the apoptotic and necrotic genetic networks. Unexpectedly, the gain of function of BCL2, which leads to a null probability of reaching apoptosis, together with the gain of function of BAX increases the apoptotic probability of BAX gain of function alone. In fact, BCL2 gain of function is able to block very efficiently both apoptosis and necrosis. If BAX gain of function promotes apoptosis as observed experimentally, deleting any signal from the necrotic (or necroptotic) pathway seems to increase apoptosis even more. This observation confirms the mutual exclusive nature of the two phenotypes. In accordance with Drees et al. \cite{Drees2005}, this type of single-nonmonotonic interactions occur with a high frequency in our networks but also in experimental data even though they are not ``recognized by common genetic nomenclature".

The distribution of all single and double mutant models forms a set of points in the multi-dimensional space of model phenotype probabilities. We found it very informative to visualize this set with the projection from multi-dimensional to two-dimensional space, using advanced methods of non-linear data visualization such as the projection onto the principal manifolds constructed by the elastic maps method (Figure~\ref{cellfate_elmap}). In these visualizations, one can see that single and double mutants form clusters characterized by some typical phenotype probability values. The cluster around the wild-type model, collects those mutants whose effect can be considered as neutral. Some clusters represent the mutants with extreme effect of induction of some of the phenotypes. The probability of different phenotypes changes along the non-linear directions (gradients) of increasing phenotype probability. Some clusters, labelled here by ``BCL2\_oe" and ``ROS\_ko" single mutants, correspond to some particular states of the model (``naive survival" for ``BCL2\_oe" and a complex state combining apoptosis and ``naive survival" for ``ROS\_ko": note that the last state can be artificial due to some irrealistic assumptions such as non-production of ROS, over-abundance of ATP, or impossibility of MPT).

\subsection{MAPK model}

The MAPK pathway controls several cellular processes such as cell cycle activation, apoptosis, survival or differentiation. The model of Grieco et al.\cite{Grieco2013} details the crosstalk between the pathways of the three mitogen-activated protein kinases: ERK, JNK and p38. In response to four stimuli (EGFR, FGFR3, TGFbeta, and DNA damage), the model produces in silico the cell response in terms of proliferation, growth arrest and apoptosis in diverse conditions, and simulates different sets of mutations often found in cancer. Even though the model is generic, its analysis is applied to studying bladder cancerogenesis.

\begin{figure}
\centering
  \includegraphics[width=8cm]{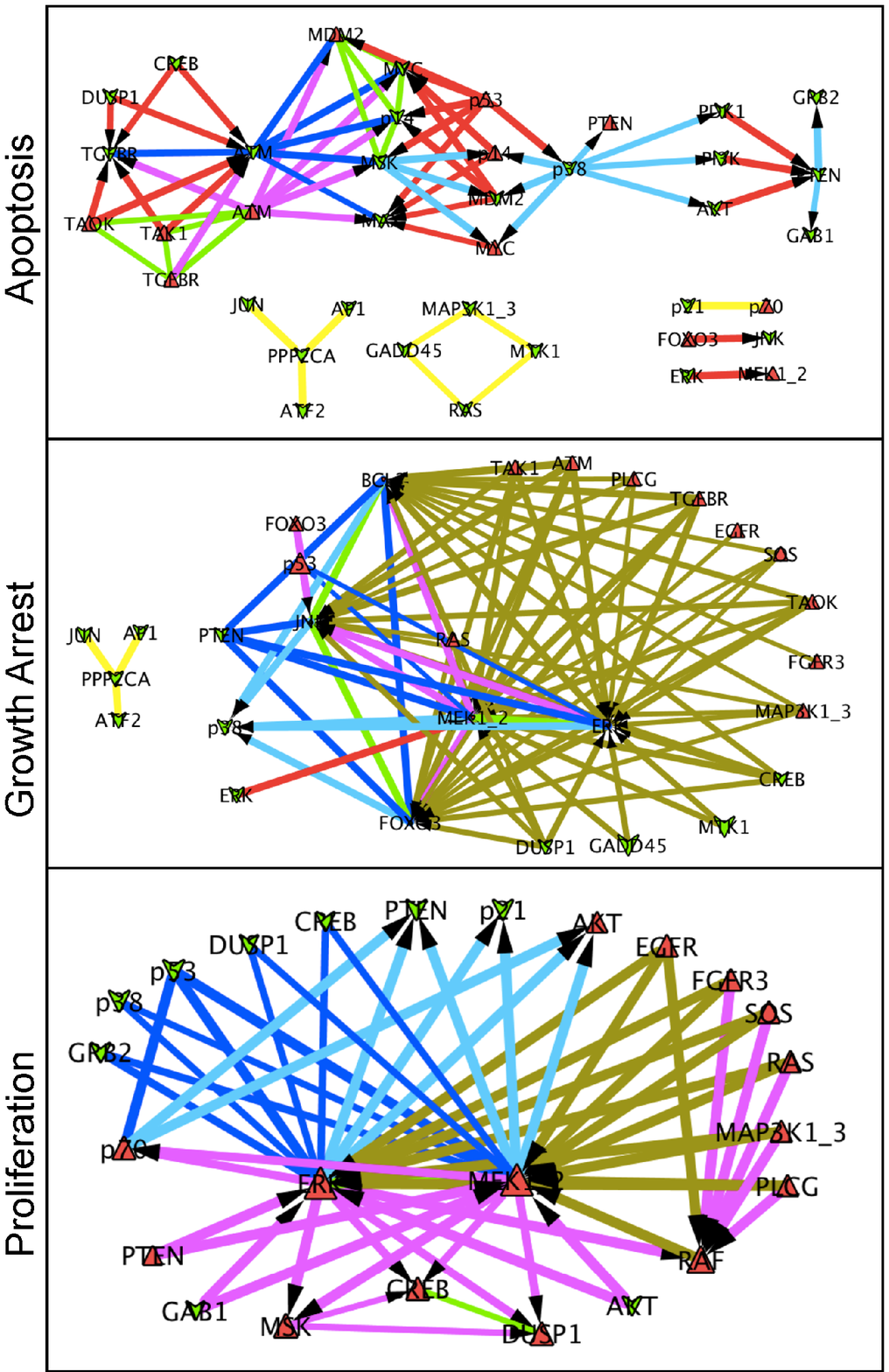}
  \caption{Genetic interaction networks computed for MAPK model, with random initial conditions and for the three phenotypes: apoptosis, growth arrest and proliferation}
  \label{MAPK_gin}
\end{figure}

Three GINs are generated using stringent conditions (interactions are selected above $k=3$ standard deviations) for filtering the edges for the three phenotypes: apoptosis, growth arrest and proliferation. The networks are characterized by modular structure, in particular, for the apoptotic phenotype (Figure~\ref{MAPK_gin}, panel 1). Interestingly, interactions within some modules or between modules are monochromatic with respect to the type of the genetic interactions. For example, a module connecting several transcription factors (JUN, AP1, ATF2) with phosphatase PPP2CA negatively controlling cell growth appears in the GINs for both the apoptosis and growth arrest phenotypes. All interactions inside this module are of ``synthetic" type (i.e., synergistic). Monochromatic structure of interactions between modules can be seen in Figure~\ref{MAPK_gin}, panel 3, where the network can be decomposed into several modules (e.g., PTEN/p21/AKT versus p70/ERK/MEK1\_2) based on the same type of interactions in between them.

Genes of the apoptotic pathway such as ATM, MAX, etc. appear to be hubs in the network with the emphasis on ATM and conditions for the two following situations: loss of function or gain of function of ATM and the partners that contribute to increasing (or compensating for the loss of) apoptosis (Figure~\ref{MAPK_gin}, panel 1). The combination of p53 gain of function and ERK gain of function seems to be a good combination to improve the growth arrest phenotype (Figure~\ref{MAPK_gin}, panel 2) whereas loss of function of PTEN reduces the arrest caused by gain of function of BCL2. In the GIN for the proliferation phenotype (Figure~\ref{MAPK_gin}, panel 3), the gain of function of either MEK1\_2 or ERK seems to be crucial in promoting proliferation, particularly in combination with gain of function of AKT or loss of function of p53 or p38, for instance. They form a hub in the network and seem to be very similar (symmetric) in terms of genetic interactions they share with the rest of the proteins of the network.

The MAPK model is the biggest network we study here. We anticipate that in even bigger regulatory network models, the corresponding genetic interaction networks should be modular and provide informative hints on pathways that are activated with respect to a particular phenotype. Predictions about the co-occurrence or the mutual exclusivity between gene alterations could be also derived from these networks.

\subsection{Mammalian restriction point model}

This Boolean model \cite{Faure2006} was adapted from a mathematical model based on ordinary differential equations developed by Novak and Tyson\cite{Novak2004}. The model was built to illustrate the behaviour of cells exposed to cycloheximide treatments at different times of the cell cycle. The model describes the dynamics of the restriction point situated in late G1 after which the cell commits to division even if treated by the drug.

\begin{figure}
\centering
  \includegraphics[width=8cm]{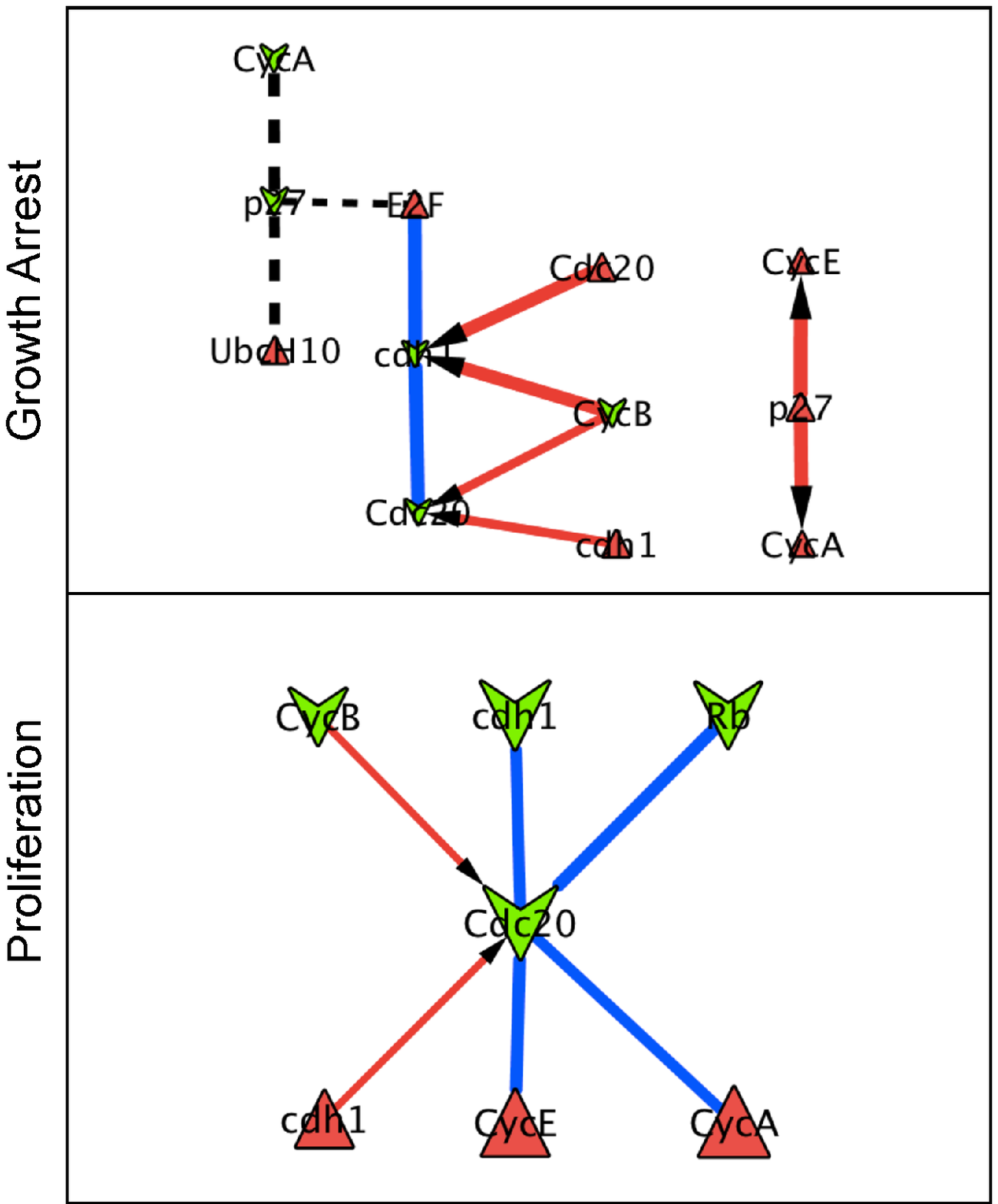}
  \caption{Genetic interaction networks computed for restriction point model, with random initial conditions and for the two phenotypes: growth arrest and proliferation}
  \label{restriction_gin}
\end{figure}

For this small model, the GINs are easier to interpret biologically (Figure~\ref{restriction_gin}). The model is built such that if the cell does not receive any external growth signals, of which CycD is the sensor, it remains stuck in G1 cell cycle phase. Therefore, neither CycD nor Rb are included in these networks as their gain or loss of function would automatically lead to forcing or deleting the phenotypes. The gain of function of the cell cycle inhibitor p27 is counteracted by the gain of function of downstream cyclins such as CycA and CycE. Similarly, if both inhibitors of the G2 and M cyclins are deleted, Cdc20 and cdh1, it is equivalent to overexpressing the cyclins and the cells can no longer arrest. A similar mechanism is achieved by overexpressing E2F and deleting cdh1. The role of cdh1 seems to be more prevalent in degrading the cyclins. Note that cdh1 and Cdc20 are in both genetic interaction networks for growth arrest and proliferation because the two read-outs are symmetric.  The loss of function of both Cdc20 and cdh1 leads to a very low probability of arresting the cycle, and a very high probability for proliferating. The two phenotypes are mutually exclusive.

\subsection{Comparison with experimentally derived genetic interactions}

We performed two types of comparisons: first, we compared the genetic interactions from our method to available experimental results, and second, we compared the genetic interactions between models.

We have compared the results from each of the examples we have chosen in this analysis with genetic interactions listed in BioGRID database \cite{Stark2006}. In the database, we queried the genes that appeared as participating in pairs of genetic interactions in a significant manner in our three models. We found that in the MAPK model, TP53 and MDM2 interactions came out in both BioGRID and our study: TP53 and MDM2 were identified in a phenotypic suppression type of genetic interaction in BioGRID and we showed that overexpression of both TP53 and of MDM2 led to a suppressive genetic interaction with respect to the apoptosis phenotype. The pair ATM and TP53 seems to be involved in a phenotypic enhancement in BioGRID, but was not found in our study. In the cell fate model, we listed three phenotypic suppressions between XIAP and CASP8, IKK1 and TNF, and BCL2 and CASP8. The first two were confirmed in our analysis: overexpression of XIAP and of CASP8 lead to an epistatic interaction with respect to apoptosis in the TNF-activated signal, and deletion of IKK1 and deletion of TNF lead to an epistatic interaction with respect to the necrosis (NonACD) phenotype in the TNF-activated signal. Also, overexpression of IKK1 and deletion of TNF lead to an epistatic interaction with respect to the survival phenotype in the TNF-activated signal. The last interaction was not identified with our method. In the mammalian restriction point model, there was only one interaction that appeared in BioGRID and involved a phenotypic enhancement between p21 and p27 which was not found in our analysis. More details can be found in supplementary materials, SuppMat\_Analysis\_BioGRID. In conclusion, the comparison showed that some interactions predicted by our method were indeed confirmed in BioGRID database. This type of comparison can serve to validate Boolean models developed for various molecular mechanisms with respect to known genetic interactions and provide additional constraints on the choice of model network topology, logical rules and rate parameters. Of course, in this analysis one should take into account incompleteness of our knowledge on genetic interactions.

We also compared more in detail the results of the genetic interactions among the three examples. Unfortunately, there was no overlap between the three models since the only common gene was BCL2 between the cell fate and the MAPK models. We then looked more carefully at the genetic interactions between phenotypes but for each model individually. With this comparison, we identified the complementary role of some genes in the networks and confirmed findings from the initial publications. The results can be found in supplementary materials, SuppMat\_comparison\_phenotypes.

\section{Conclusions}

In this manuscript, we suggest a methodology for converting a logical mathematical model with a set of initial conditions into the corresponding genetic interaction network characterizing the behaviour of all single and double mutants in terms of phenotype probabilities. The advantage of the methodology is in that it allows:

1) estimating and classifying possible functional interactions between the different elements composing the model;

2) distinguishing extreme cases of mutations amplifying or masking each other and, based on this, suggesting intervention points in order to achieve a desired phenotype (such as in \cite{Chanrion2014});

3) suggesting experimental designs from the logical models;

4) detecting controversial (non-intuitive) properties of mutants with respect to expected phenotypes such as nonmonotonic genetic interactions;

5) comparing quickly similar logical network models in terms of their functional properties;

6) validating the model and comparing different models using available screenings for genetic interactions (such as synthetic lethality screens).

The last point deserves further development. We aim at extending our methodology using existing databases containing genetic interactions (similar to what we did with BioGRID \cite{Stark2006}) for matching the model predictions with genetic interactions or 
 single mutation phenotypes known from the literature or from screenings. Moreover, similar to the methodology of parameter fitting in constructing chemical kinetic models, one can fit the kinetic rates defined in our continuous-time discrete approach \cite{Stoll2012} in order to optimize the set of model predictions. Another set of experimental data that could be used with this approach is high-throughput cancer data, such as large-scale mutation landscapes that are collected for series of tumours. Patterns of co-occurrence or mutual exclusivity of mutations can reflect action of genetic interactions in cancer cells. For example, synthetically lethal interactions can lead to the pattern of mutual exclusivity since cancer cells possessing both synthetically lethal mutations will be eliminated from the cell population. Using these data for interpretation and validation of model-based predictions requires the development of a statistical methodology for detecting statistical patterns in high-throughput data.

Genetic interaction networks reconstructed from logical mathematical models possess many properties of experimentally-measured networks. They are characterized by a variety of types of genetic interactions (with predominance of masking, e.g., epistatic interactions), modular structure for sufficiently big discrete models (Figure~\ref{MAPK_gin}), with some modules characterized by monochromaticity for within-module interactions as well as between-module interactions. Sets of genetic interactions are highly dependent on the phenotype with respect to which they are defined and, to less extent, sensitive to the initial conditions (in other words, to the molecular context) chosen for performing simulations. These properties make the obtained genetic interaction networks a good model for the experimentally-measured ones.

Therefore, we believe that the suggested methodology will contribute to the toolbox of computational approaches in systems biology, connected to mathematical modeling of cellular mechanisms.


\section{Acknowledgements}

This work was supported by internal project of Institut Curie ``PIC Computational Systems Biology of Cancer".

\bibliography{Calzone2015IntegBio} 
\bibliographystyle{natbib} 

\end{document}